# Performance Evaluation of TCP over Mobile Ad-hoc Networks

Foez ahmed[1], Sateesh Kumar Pradhan[1], Nayeema Islam[2], and Sumon Kumar Debnath[3]

[1]Dept. of Computer Networks Engineering, College of Computer Science, King Khalid University, Kingdom of Saudi Arabia
[2]Department of Information & Communication Engineering, Rajshahi University, Rajshahi- 6205, Bangladesh
[3]Dept. of Computer Science and Telecommunication Engineering, Noakhali Science and Technology University, Bangladesh

*Abstract*— With the proliferation of mobile computing devices, the demand for continuous network connectivity regardless of physical location has spurred interest in the use of mobile ad hoc networks. Since Transmission Control Protocol (TCP) is the standard network protocol for communication in the internet, any wireless network with Internet service need to be compatible with TCP. TCP is tuned to perform well in traditional wired networks, where packet losses occur mostly because of congestion. However, TCP connections in Ad-hoc mobile networks are plagued by problems such as high bit error rates, frequent route changes, multi-path routing and temporary network partitions. The throughput of TCP over such connection is not satisfactory, because TCP misinterprets the packet loss or delay as congestion and invokes congestion control and avoidance algorithm. In this research, the performance of TCP in Ad-hoc mobile network with high Bit Error rate (BER) and mobility is studied and investigated. Simulation model is implemented and experiments are performed using the Network Simulatior-2 (NS-2).

## I. INTRODUCTION

An ad hoc network is a collection of mobile nodes forming a temporary wireless network without the aid of any established infrastructure or centralized administration [1]. Since each node in the network is potentially mobile, the topology of an ad-hoc network can be highly dynamic. Ad hoc networks are very useful in emergency search-and-rescue operations, meetings or conventions in which persons wish to quickly share information, and data acquisition operations in inhospitable terrain.

In ad hoc networks all nodes are mobile and can be connected dynamically in an arbitrary manner. All nodes of these networks behave as routers and take part in discovery and maintenance of routes to other nodes in the network. This ad-hoc routing protocols can be divided into two categories: Table-driven routing protocol and On-Demand routing protocols. In table driven routing protocols, consistent and up-to-date routing information of all nodes is maintained at each node. In On-Demand routing protocols, the routes are created when required. When a source wants to send data to a destination, it invokes the route discovery mechanisms to find the path to the destination. In recent years, a variety of new routing protocols targeted specifically at this environment have been developed. There are four multi-hop wireless ad hoc network routing protocols that cover a range of design choices: Destination-Sequenced Distance-Vector (DSDV) [2], Dynamic Source Routing (DSR) [1], On-Demand Distance Vector Routing (AODV) [3] and Temporally Ordered Routing Algorithm (TORA) [4]. While DSDV is a table-driven routing protocol, TORA, DSR, AODV, fall under the On-demand routing protocols category.

As the standard network protocol on the Internet, the use of the TCP/IP over the ad-hoc network is a certainty. It is because that the use of TCP/IP supports a large number of popular applications and allows seamless integration with the internet. In addition, TCP/IP provides network-level consistency for multi-hop networks that consist of hosts using a variety of physical-layer media. However, TCP assumes a relatively reliable underlying network where most packet losses are due to congestion, which conflicts strongly with characteristics of wireless links in ad-hoc mobile network. In mobile ad hoc environment losses are more often caused by high Bit Error Rate (BER), route re-computation, temporary network partition, and multi-path routing [5]. If the losses are not due to congestion, then TCP unnecessarily reduces throughput leading to poor performance.

In this paper, the effect of high bit error rate and route re-computation on the performance of TCP in mobile ad hoc network is analyzed. All four mentioned routing protocols are used throughout the experimentations. The TCP variants used in these simulations are TCP Tahoe [6], Reno [7], New Reno [8] and Sack [9].

The rest of this paper is organized as follows. In section II the impact of multi-hop wireless networks on TCP performance is discussed briefly. The network topologies together with simulation parameters are presented in section III. Simulation results are also discussed in the same section. Finally, the recommendations for future work in this area and concluding remarks are provided in section IV.

## II. THE PROBLEMS WITH TCP IN AD HOC NETWORKS

This section describes the effects of High Bit Error Rate and Route re-computation on the performance of TCP and the possible reasons behind these effects.

### A. Effect of High Bit Error Rate (BER)

Bit errors cause packets to get corrupted resulting in lost TCP data segments or acknowledgements. When acknowledgements do not arrive at the TCP sender within a given short amount of time (the retransmit timeout or RTO), the sender retransmits the segment, exponentially backs off its retransmit timer for the next retransmission, and decreases its congestion window. Repeated errors result in a small congestion window at the sender and causes low throughput.

### B. Effect of Route Re-computations

When an old route is no longer available, the routing protocol at the sender attempts to find a new route to the destination. It is possible that discovering a new route may take long. As a result, the TCP sender times out, retransmits a packet, and invokes congestion control mechanisms. This is clearly an undesirable behavior because the TCP connection will be very inefficient. If we imagine a network in which route computations are done frequently (due to high node mobility), the TCP connection will never get an opportunity to transmit at the maximum negotiated rate (i.e., the congestion window will always be significantly smaller than the advertised window size from the receiver).

## III. EXPERIMENTAL SETUP, SIMULATION RESULTS AND DISCUSSIONS

### A. Measurement parameter

The performance metrics of the interest in this study is throughput, which is the measure of how soon an end user is able to receive data. It is determined as the ratio of the total data received by the end user and the connection time. A higher throughput will directly impact the user's perception of the quality of service.

### B. Simulation Tools

The results in this study are based on simulations using the network simulator (NS2) from Lawrence Berkeley National Laboratory (LBNL) , with extensions from the MONARCH project at Carnegie Mellon [10]. The extensions include a set of mobile ad hoc routing protocols as well as an 802.11 MAC layer and a radio propagation model.

### C. Effect of High Bit Error Rate

To investigate the effect of high BER on the TCP performance two different types of topologies are considered: Grid topology and Chain topology. Both topologies are considered to be static.

### C.1. Experimental setup for Chain Topology and Result analysis

A 7-node chain or string topology is shown in fig. 1. In a chain topology, TCP packets travel along a chain of intermediate nodes toward the destination.

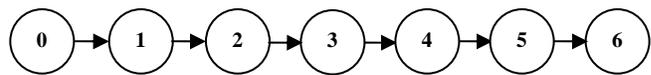

Figure 1. Chain Topology

Simulations are performed to investigate the performance of the TCP flow whose source and destination are placed at both ends of the chain topology. A single wireless channel is shared for transmissions, and only receivers within the transmission range of the sender can receive the packets. The two adjacent nodes are about 200m apart. In this simulation the data rate of the wireless channel is the two-ray ground model with transmission range 250m, carrier sensing range 550m, and interference range 550m. The network model consists of 4, 7 and 10 nodes respectively in a 2200*500 meter flat, rectangular area. At first simulations are performed without introducing any error on the links. Throughputs (in Mbps) of various TCP variants which are TCP Tahoe, TCP Reno, TCP New Reno and Sack over mobile ad hoc networks are identified. For performance comparison all four routing protocols i.e., DSDV, AODV, DSR, TORA are considered. These results for 4-node, 7-node and 10 node networks are shown in table I.

TABLE I. THROUGHPUT OF TCP VERSIONS FOR CHAIN TOPOLOGY

| | 4-Node Chain Topology | | | | 7-Node Chain Topology | | | | 10-Node Chain Topology | | | |
|---|---|---|---|---|---|---|---|---|---|---|---|---|
| | DSDV | AODV | DSR | TORA | DSDV | AODV | DSR | TORA | DSDV | AODV | DSR | TORA |
| Tahoe | 1.248 | 1.603 | 1.603 | 1.602 | 1.248 | 1.603 | 1.603 | 1.602 | 1.248 | 1.603 | 1.603 | 1.602 |
| Reno | 1.248 | 1.603 | 1.603 | 1.602 | 1.248 | 1.603 | 1.603 | 1.602 | 1.248 | 1.603 | 1.603 | 1.602 |
| New Reno | 1.313 | 1.686 | 1.685 | 1.685 | 1.318 | 1.686 | 1.686 | 1.685 | 1.313 | 1.686 | 1.686 | 1.685 |
| Sack | 1.248 | 1.603 | 1.603 | 1.602 | 1.248 | 1.603 | 1.603 | 1.602 | 1.248 | 1.603 | 1.603 | 1.602 |

From the table I, it is observed that maximum throughput is achieved for New Reno protocol for all four ad hoc routing protocols. However if routing protocols are compared, AODV shows the best performance. From these results it can be concluded that New Reno-AODV combination shows the best performance for all the sizes of chain topologies. Hence this combination is used to identify the effect of high BER. Simulations are performed to investigate the performance of TCP for a variety of wireless packet error rates ranging from 0.000 to 0.017. Each simulation is performed for 200 seconds.

The observed throughputs for all the networks are plotted in fig. 2. From figure it can be observed that TCP throughput highly degrades with increasing packet error rate that proves its poor performance. Since TCP's congestion control algorithm is responsible for TCP throughput it leads to poor performance in High error rate environment. This figure also compares the impact of varying network sizes in presence of Bit Error. It can be observed that as the network size increased, throughput tends to decrease earlier with increasing PER. For example when the packet error rate is 0.015 the throughput for 4 node chain topology is 0.0296559 Mbps, the throughput of 7 node chain topology is 0.001024 Mbps and finally the throughput of 10 node chain topology is 0.00Mbps. Hence, not only the increasing bit error rates effects the network throughput but also the increasing network size have significant effect on TCP performance.

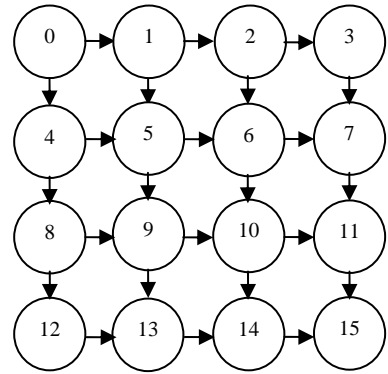

Figure 3. 16-node Grid Topology

The performance of the TCP flow is investigated whose source and destination is placed at two opposite corners of the grid (Bottom-left and Upper-Top of the grid). For the 16-node network the distance between two adjacent nodes is 100 meter. For other two networks the nodes are 50 meters apart. The data rate of the wireless channel is the two-ray ground model with transmission range 250m, carrier sensing range 550m, and interference range 550m. The size of all networks is 500m × 500m. . Again simulations are performed without introducing any error on the links. Throughputs (in Mbps) of all TCP variants all four mentioned routing protocols are calculated. These results are shown in table II.

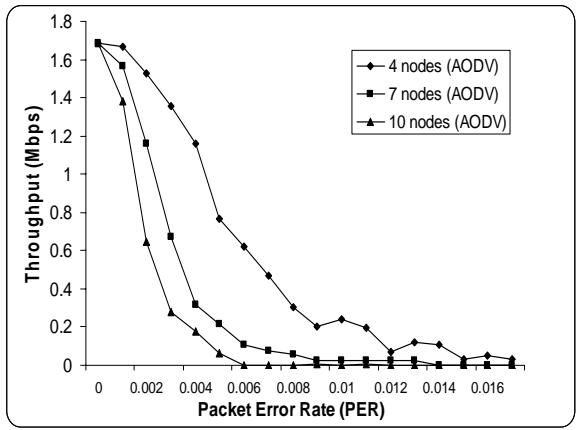

Figure 2. Throughput of TCP New Reno over ad hoc network for PER ranging from 0.0 to 0.016 for various sizes of Chain Topologies.

TABLE II. THROUGHPUT OF TCP VERSIONS FOR GRID TOPOLOGY

|  | 16-Node Grid Topology | | | | 25-Node Grid Topology | | | | 49-Node Grid Topology | | | |
| --- | --- | --- | --- | --- | --- | --- | --- | --- | --- | --- | --- | --- |
|  | *DSDV* | *AODV* | *DSR* | *TORA* | *DSDV* | *AODV* | *DSR* | *TORA* | *DSDV* | *AODV* | *DSR* | *TORA* |
| Tahoe | 1.4995 | 1.6029 | 1.6024 | 1.5938 | 1.2996 | 1.6029 | 1.6024 | 1.5937 | 1.5677 | 1.6029 | 1.7795 | 1.5929 |
| Reno | 1.4981 | 1.6029 | 1.6024 | 1.5945 | 1.2964 | 1.6029 | 1.6024 | 1.5944 | 1.5666 | 1.6029 | 1.7795 | 1.5936 |
| New Reno | 1.5492 | 1.6859 | 1.6853 | 1.6769 | 1.3313 | 1.6859 | 1.6854 | 1.6768 | 1.6310 | 1.6859 | 1.7786 | 1.6760 |
| Sack | 1.4980 | 1.6029 | 1.6024 | 1.5938 | 1.3001 | 1.6029 | 1.6024 | 1.5937 | 1.5667 | 1.6029 | 1.7795 | 1.5929 |

*C.2. Experimental setup for Grid Topology and Result analysis*

Three different sizes of Grid topologies are used for simulation. The numbers of nodes of these grid topologies are 16, 25, and 49 respectively. A 16-node grid topology is shown in fig. 3.

For 16-node grid topology New Reno achieves maximum throughput for all the ad hoc routing protocols. AODV routing protocol gives the best performance when New Reno is chosen as the transport layer protocol. Same result is obtained for 25 node grid topology. In case of 49 node grid topology New Reno shows the best performance. However in this case DSR is the dominant routing protocol since it helps to achieve maximum throughput.

Next, errors are generated for all the networks. The packet error rate is varied from 0.000 to 0.017. The calculated throughputs are plotted in fig. 4. From figure 4 can be observed that increasing error decreases the TCP throughput for all network sizes. For 25-node network, throughput decreases rapidly than 16-node network with increasing PER. For 49-node network throughput decreases slowly than other two. Again, the reduction of throughput of TCP is due to mistaking wireless error as an indication of congestion.

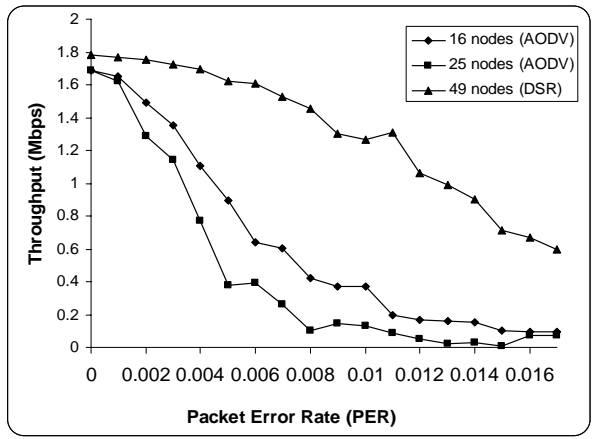

Figure 3.  Throughput of TCP New Reno over ad-hoc network for PER ranging from 0.000 to 0.016 for various sizes of grid topologies

### D. Effect of Route Re-computation on the Performance of TCP

The effects of route re-computation are investigated for AODV and TORA routing protocols.

#### D.1. Effect of Route Re-computation for AODV Routing Protocol

To experiment the effect of route re-computation on the performance of TCP the following network topology (fig. 5) is used.

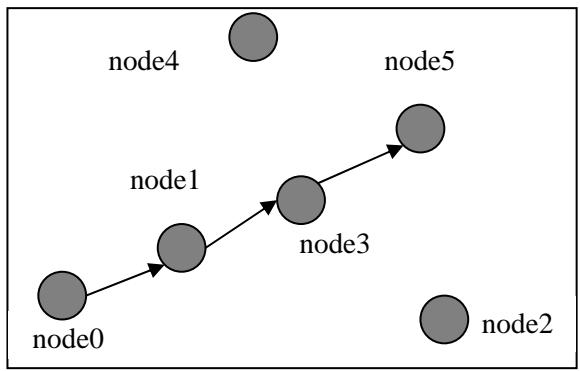

Figure 4.  Network Topology before Mobility (AODV Routing Protocol)

There are 6 nodes in this network. The scenario area is 600 m × 600 m and transmission Range of each node is 250 m. All nodes communicate with identical wireless radios, which have a bandwidth of 2Mbps. It is assumed that all links are error free. New Reno is used as the TCP standard. The simulation is run for a period of 200 seconds. The initial positions of the nodes are, node0: (25, 25), node1: (156,143), node2: (396,143), node3: (303,242), node4: (160,331), node5: (399,329).

During simulation node0 starts sending packets at 5 seconds to node5 through the route node0→node1→node3→node5 (figure 5). At time 10 seconds node3 starts moving towards the coordinate position (599,599) at the speed of 30m/s. At the same time node2 also start moving towards the coordination position (303,242) at the same speed as the node3. Finally a new network topology is formed as shown in fig. 6.

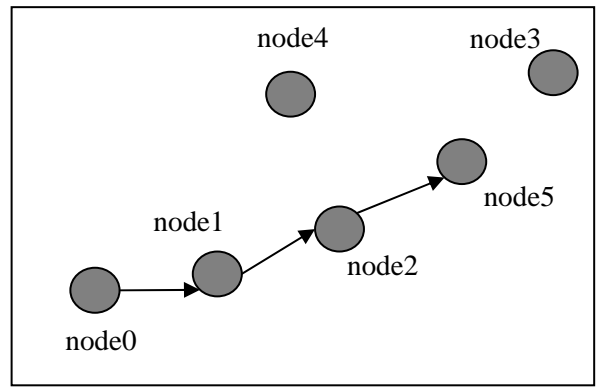

Figure 5.  Changed Network Topology after Mobility (AODV Routing Protocol)

Due to mobility node3 moves out of the transmission range and within few seconds the transmission link breaks. Now the old route is no longer available and hence the routing protocol at the sender tries to find a new route to the destination. From the simulation scenario it can be observed that if all other nodes are fixed it is not possible for the sender to find out the new route due the unavailability of the wireless transmission range. To establish a connection a node must move towards the transmission range, which we generate through the movement of node2. At time nearly 40 seconds node2 reaches the transmission range. At the same time a new route is established and node0 again starts sending packets through the new route node0→node1→node2→node5. It is seen that discovering a new route takes about 35 seconds. Meanwhile the TCP sender has timed out and has invoked congestion control mechanism, which results sever degradation of TCP throughput. Fig. 7 shows the performances of this network for the entire simulation time. From the graph it can be observed that during time interval 10 to 40 seconds no packet is delivered from the source to the destination. Hence the throughput remains zero for that time period. After 40 seconds the connection resumes due to the new route and throughput begins to increase.

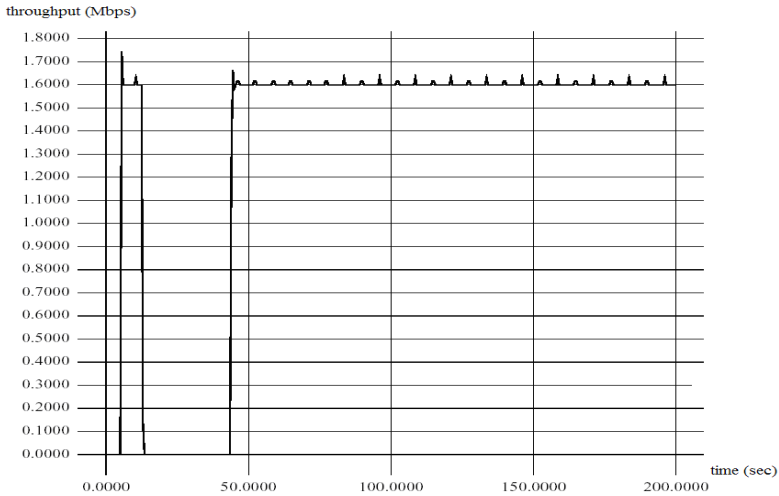

Figure 6.  Throughput of TCP over Mobile Ad-hoc network for the entire Simulation time of 200 seconds in presence of Dynamic Network Topology (AODV Routing Protocol).

### D.2. Effect of Route Re-computation for TORA Routing Protocol

To investigate the effect of route re-computation on the performance of TCP in case of TORA network protocol the same topology is used as AODV (Figure 5). The network parameters are also same. In this case the sender node0 has established the route through node1 and node2 to send packets to node5. The route is shown in fig. 8

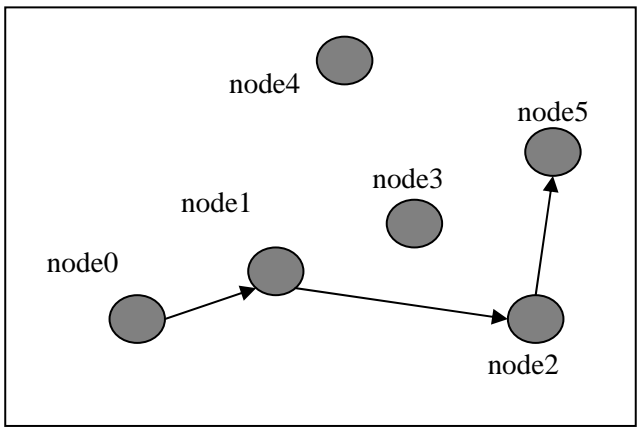

Figure 7.  Network Topology before Mobility (TORA Routing Protocol)

At time 10 seconds node2 starts moving towards the coordinate position (599,599) at the speed of 30m/s. When this node moves out of the wireless transmission range the connection breaks and packets start dropping. TCP assumes these packet drops as the indication of congestion and invokes congestion control algorithm. Meanwhile node0 tries to find out a new route and at near about 20 seconds it established a new route through the nodes1 and node3 and again resumes sending packets. The new route is shown in fig. 9.

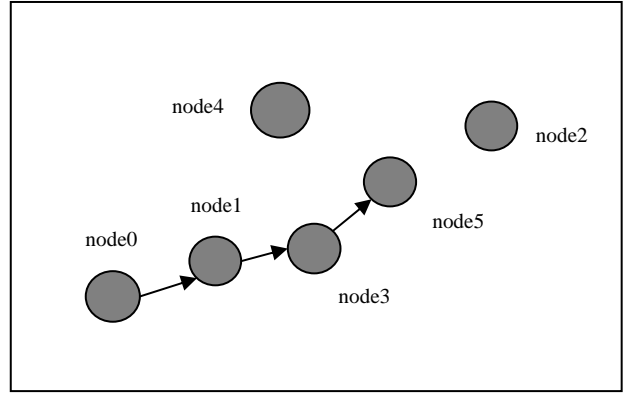

Figure 8.  Changed Network Topology after Mobility (TORA Routing Protocol)

The TCP throughput for the entire simulation time is shown in the graph of figure 10. From the graph it is clear that between the time intervals 10 to 20 seconds no packet is successfully delivered to the destination and hence the throughput reduces zero. After 20 seconds when the connection is re-established throughput starts increasing and remains almost same for the remaining time.

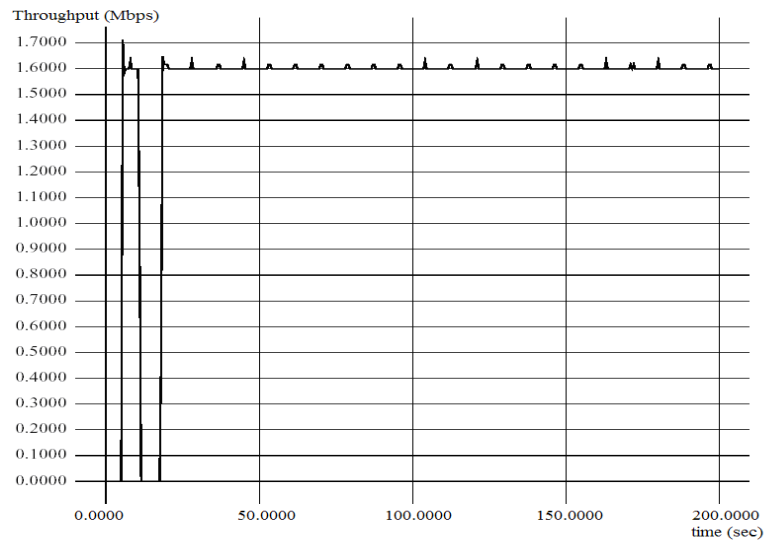

Figure 9.  Throughput of TCP over Mobile Ad-hoc network for the entire Simulation time of 200 seconds in presence of Dynamics Network Topology (TORA Routing Protocol)

## IV. Conclusions And Recommendations

In this paper, TCP performance is evaluated in a multi-hop wireless network environment. The impact of high bit error rate of wireless links on the performance of TCP is investigated through simulation for both the chain and grid topologies. TCP cannot differentiate packet lost due to congestion and packet lost due to bit error. For both cases, the TCP sender assumes that networks congestion has occurred and invokes congestion control algorithm. From simulations it is observed that this algorithm results in significant reduction in throughput and unacceptable inter-reactive delays for active connections, thus severely degrading performance. Due to dynamic topology of ad hoc network the route between a sender and receiver changes rapidly. From simulation it is seen that the route change results in link disconnections, which reduces TCP throughput. Also sometimes it takes long to find a new route to resume data transfer. Meanwhile TCP sender times out and invokes congestion control mechanisms. From experiments it is also observed that due to difference of routing protocols, same topology uses different routes to transmit data.

Further investigation can be carried out to study the effects of network partitions and multi-path routing on the TCP performance. The impact that the link-layer has on TCP performance, such as aggregate delay caused by local retransmissions over multiple wireless hops can be experimented through simulation. Experiments can be carried out to evaluate the performance in a congested ad hoc network.

Authors Profile

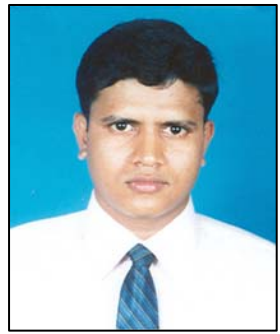

**Foez Ahmed** joined as a Lecturer in the Dept. of Electronics and Communication Engineering, Northern University Bangladesh (NUB), Dhaka, Bangladesh in the 2008. At present he is on leave from Northern University and working as a Lecturer with the Dept of Networks and Communication Engineering, College of Computer Science, King Khalid University, Kingdom of Saudi Arabia. He did his under graduation and post graduation in Information and Communication Engineering in 2007 and 2009 respectively from Rajshahi University, Bangladesh. He has received various Awards and Scholarships for the under graduation and post graduation results. His research interests include Mobile Ad-hoc Networks and Routings, Cognitive Radio Networks, Cooperative Communications, Sensor Networks and Sparse Signal Processing in Wireless Communication.

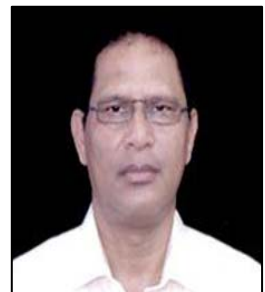

**Sateesh Kumar Pradhan** obtained his PhD degree in Computer Science from Berhampur University, India during the year 1999. He joined Berhampur University, as Assistant Professor in the 1987 and promoted to Associate Professor in 1999. He was Head of the Department of Computer Science, Utkal University, India during 2001-2003. He was the Secretary and Vice-President Orissa IT Society from 2003-2005 and 2005-2007 respectively and was the organizing Chair of the International Conference on Information Technology – 2005 (ICIT – 2005). At present he is on leave from Utkal University and working with the Department of Computer Engineering, King Khalid University, KSA. His research interests include Neuron based Algorithms, Computer Architecture, Ad-hoc Network and Computer Forensic.

**Nayeema Islam** was born 1976 in Bangladesh. She received her M.S. in Telecommunication Engineering from Asian Institute of Technology, Thailand in 2004. Also she received her M.Sc. and B.Sc. in Computer Science and Technology from Rajshahi University, Bangladesh in 1998 and 1997 respectively. She is presently working as Assistant Professor in the department of Information and Communication Engineering, University of Rajshahi since 2004. Her fields of interest include Telecommunications, computer networking, mobile ad-hoc networks and QoS routing.

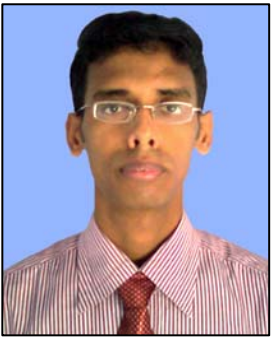

**Sumon Kumar Debnath** obtained his M.Sc degree in Information and Communication Engineering from Rajshahi University, Bangladesh during the year 2009. He has joined Noakhali Science and Technology University, Bangladesh and working as a lecturer in the department of Computer Science and Telecommunication Engineering. He is also working as a Assistant Proctor in that university. His research interests include Mobile Ad-hoc network, Sensor Network, MIMO, OFDM, and MCCDMA.